\documentclass[reprint,
 amsmath,amssymb,aps,,superscriptaddress]{revtex4-1}
\usepackage{color}
\usepackage[toc,page]{appendix}
\usepackage[english]{babel}
\usepackage{microtype}
\usepackage{braket}
\usepackage{graphicx}
\usepackage{dcolumn}
\usepackage{bm}
\usepackage[normalem]{ulem}
\usepackage[colorlinks]{hyperref}
\hypersetup{
    colorlinks=true,       
    linkcolor=blue,         
    citecolor=blue,        
    filecolor=blue,     
   urlcolor=blue         
}

\begin{document}

\title{The Hofstadter Butterfly in a Cavity-Induced Dynamic  Synthetic Magnetic Field}

\author{Elvia Colella}
\email{elvia.colella@uibk.ac.at}
\affiliation{Institut f\"ur Theoretische Physik, Universit{\"a}t Innsbruck, A-6020~Innsbruck, Austria}
\author{Farokh Mivehvar}
\affiliation{Institut f\"ur Theoretische Physik, Universit{\"a}t Innsbruck, A-6020~Innsbruck, Austria}
\author{Francesco Piazza}
\affiliation{Max-Planck-Institut f\"{u}r Physik komplexer Systeme, D-01187 Dresden, Germany}
\author{Helmut Ritsch}
\affiliation{Institut f\"ur Theoretische Physik, Universit{\"a}t Innsbruck, A-6020~Innsbruck, Austria}

\begin{abstract}
Energy bands of electrons in a square lattice potential threaded by a uniform magnetic field exhibit a fractal structure known as the Hofstadter butterfly. Here we study a Fermi gas in a 2D optical lattice within a linear cavity with a tilt along the cavity axis. The hopping along the cavity axis is only induced by resonant Raman scattering of transverse pump light into a standing wave cavity mode. Choosing a suitable pump geometry allows to realize the Hofstadter-Harper model with a cavity-induced dynamical synthetic magnetic field, which appears at the onset of the superradiant phase transition. The dynamical nature of this cavity-induced synthetic magnetic field arises from the delicate interplay between collective superradiant scattering and the underlying fractal band structure. Using a sixth-order expansion of the free energy as function of the order parameter and by numerical simulations we show that at low magnetic fluxes the superradiant ordering phase transition is first order, while it becomes second order for higher flux. The dynamic nature of the magnetic field induces a non-trivial deformation of the Hofstadter butterfly in the superradiant phase. At strong pump far above the self-ordering threshold we recover the Hofstadter butterfly one would obtain in a static magnetic field.   
\end{abstract}

\pacs{Valid PACS appear here}
\maketitle                        

\section{\label{sec:level}Introduction}
In the last decade, advancements in the manipulation of cold atomic gases enabled to engineer Hamiltonians emulating the physics of effective gauge fields ~\cite{goldman2014light,dalibard2011colloquium}. The development of rotating traps~\cite{engels2003observation,schweikhard2004rapidly} allowed to overcome the challenge of coupling the external degrees of freedom of neutral atoms to an effective vector gauge potential as for charged particles. More sophisticated techniques based on light-matter interaction~\cite{lin2009bose,chen2011controlling} and lattice shaking~\cite{hauke2012non, goldman2014periodically} were also developed to imprint a position-dependent geometric phase onto the atomic wave-function, analogous to the Aranov-Bohm phase of electrons in an external magnetic field~\cite{aharonov1959significance}. The Hofstadter model~\cite{harper1955single,hofstadter1976energy} was shortly after implemented for cold atoms in optical lattices by employing a laser-assisted tunneling scheme~\cite{jaksch2003creation,aidelburger2011experimental,osterloh2005cold}. The realization of such an artificial magnetic field in lattice geometries~\cite{aidelsburger2013realization,jotzu2014experimental} allows one to explore the realm of topological many-body states of matter~\cite{goldman2016topological,zhang2018topological,cooper2019topological}. The most notable examples include measuring the Chern number of non-trivial topological bands~\cite{aidelsburger2014measuring} and realizing the Meissener phases for neutral atoms in ladder geometries~\cite{atala2014observation}. More recently, new techniques exploiting internal degrees of freedom as synthetic dimension have been developed~\cite{boada2012quantum,celi20142014} and are candidates for the observation of the quantum Hall effect even in four dimensions~\cite{price2015four}.

The experimental realization of lattice models with effective gauge potential is of great interest for engineering synthetic gauge theories~\cite{wiese2013ultracold}. Experimental realizations so far implemented static gauge fields which can be finely tuned by varying experimental parameters, but are not dynamically affected by the atomic back-action. However, in order to simulate a genuine gauge theory, quantum matter needs to be dynamically coupled to a gauge (bosonic) field and the back-action of the matter dynamics onto the gauge field should be accounted for. A first step in this direction is to use density-dependent synthetic gauge fields ~\cite{edmonds2013simulating,keilmann2011statistically}, which  were recently observed for a BEC in a shaken optical lattice~\cite{clark2018observation,gorg2018realisation}. A $Z_2$ lattice gauge theory was also experimentally realized~\cite{barbiero2018coupling,schweizer2019floquet}. 
\par
Optomechanical systems~\cite{lauter2015pattern,walter2016classical} as well as cold atoms in optical cavities~\cite{ritsch2013cold} provide another natural route to the realization of a dynamical gauge theory in a controllable and accessible environment. This hinges on the non-linearity of these systems, where photons (phonons) feel the back-action of the atomic motion (photons). In view of the experimental realization of a strongly interacting Fermi gas coupled to a cavity~\cite{roux2019strongly} and the recent observation of a dynamical spin-orbit coupling in a BEC in a linear cavity~\cite{kroeze2018spinor,guo2019sign,kroeze2019dynamical,Mivehvar2018toolbox} , theoretical proposals~\cite{mivehvar2014synthetic,dong2014cavity,deng2014bose,mivehvar2015enhanced,ballantine2017meissner,zheng2016superradiance,halati2017cavity,sheikhan2016cavity,halati2019cavity} for dynamical gauge fields are now in reach by experiments. 

 \begin{figure}[t]
\centering
\includegraphics[width=0.5\textwidth]{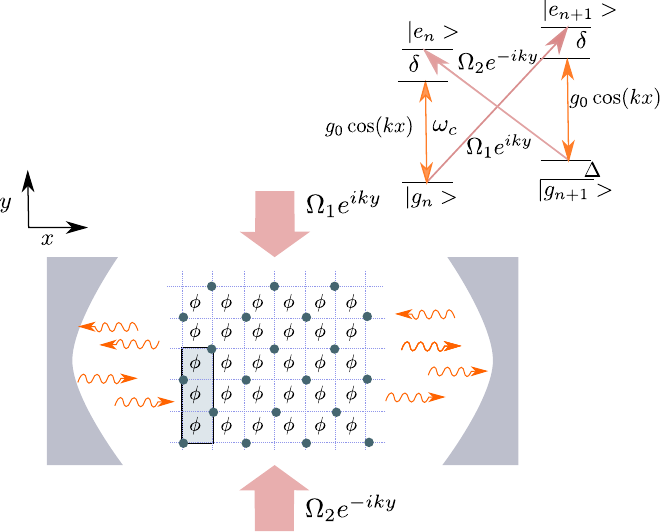}
\caption{Geometry sketch to realize a dynamical version of the Harper-Hofstadter Hamiltonian: a 2D Fermi gas in a rectangular lattice within a single-mode optical cavity is transversely illuminated by two counter-propagating laser beams of orthogonal polarization. The shaded area in the lattice represents the unit cell for $\phi=2\pi/3$.}
\label{fig:1}
\end{figure}

Here we study dynamical cavity-supported synthetic magnetic fields for fermions in an external optical lattice~\cite{jaksch2003creation}. Atoms are driven by two transverse counter-propagating lasers and can scatter photons into the cavity. The hopping along the cavity axis is suppressed by a potential gradient.  By choosing proper laser detunings, it can be activated by resonant Raman scattering of pump photons into a single resonant standing wave mode of the cavity~\cite{sheikhan2016cavity}. Each pump laser here is responsible for a particular hopping direction. Above a critical pump strength, the collective  buildup of the cavity field enables resonant coherent tunneling. In addition, for any closed loop in the atomic trajectory, a geometric phase proportional to the enclosed area is imprinted onto the atomic wave-function, in analogy to the phase acquired by electrons in a magnetic field.

The onset of the superradiant phase transition and the appearance of a synthetic magnetic field depends strongly on the phases imprinted, which can be tuned by setting the ratio between the lattice constant and the pump field wavelength $B\propto d_y/\lambda_c$. This is due to an intricate interplay between superradiant scattering generating the synthetic magnetic field and the emerging fractal energy bands corresponding to this field. Such cavity-induced atomic back-action on the effective gauge potential is very different to existing free-space implementations. Interestingly, as shown below, the onset of the superradiant phase transition (and hence appearance of the synthetic magnetic field) exhibits a first-order behavior at low fluxes, where the energy bands are Landau-like, while it becomes second-order for high flux. The energy spectrum itself carries the signs of the non-linearity of the atom-light interactions and the dynamical nature of the magnetic field, resulting in the emergence of peculiar structures compared to the commonly known energy spectrum, i.e., Hofstadter butterfly~\cite{hofstadter1976energy}.
\par
The paper is organized as follows. In section~\ref{sec:2} we introduced the detailed system model. The physical results are summarized in section~\ref{sec:3}, where we focus on the bulk properties of the system at half-filling. Here the gas behaves as a metal or semi-metal depending on the value of the magnetic flux in a plaquette. We show the phase diagram, the energy spectrum and we investigate the point of change of the phase transition from first to second order. Our final considerations are reported in section~\ref{sec:4}.

\section{\label{sec:2}Model}
We consider a Fermi gas confined in a two dimensional (2D) optical lattice of lattice constant, $\mathbf d=\{d_x,d_y\}$, in the tight binding regime. Hopping in the $x$-direction is suppressed by an additional energy gradient $\hbar\Delta$ between neighbouring sites. This can be realized by adding a constantly accelerated optical lattice, a magnetic field, or an electric field gradient along the $x$-direction. We consider only a single internal atomic transition $\ket{g}\leftrightarrow \ket{e}$ of frequency $\omega_0$. The hopping in the $x$-direction is restored via two-photon resonant scattering processes mediated by cavity photons, where the resonance condition is $\omega_c\simeq \omega_1 + \Delta=\omega_2-\Delta$~\cite{jaksch2003creation}. Here, $\omega_1$ and $\omega_2$ are the frequencies of the two transversal laser pumps; see Fig.~\ref{fig:1}. 
\par
Our model Hamiltonian in tight-binding approximation in a reference frame rotating at the average pump frequency $\omega_p=(\omega_1+\omega_2)/2$ then reads:~\cite{sheikhan2016cavity},
\begin{subequations}\label{second:main}
\begin{align}
 H=& -J_y\sum_{l,m}(f^\dagger_{l,m+1}f_{l,m}+\text{H.c.})\tag{\ref{second:main}}\\ 
  &-\hbar\eta(a+a^\dagger)\sum_{l,m}(e^{2i\pi m \gamma}f^\dagger_{l+1,m}f_{l,m}+\text{H.c.})\nonumber \\ 
  &-\hbar\Delta_c a^\dagger a.\nonumber 
\end{align}
 \label{eq:ham}%
\end{subequations}
Here $J_y$ is the hopping amplitude in the $y$-direction, $\eta=\Omega_1 g_0/\delta=\Omega_2 g_0/\delta$ is the two photon Rabi coupling with $\delta=\omega_p-\omega_0$ the atomic detuning with respect to the average pump frequency, $g_0$ is the bare coupling strength of the cavity mode to the atomic transition and $\Delta_c=\omega_p-\omega_c$ is the cavity detuning with respect to the average pump $\omega_p$. Note that only resonant Raman scattering terms are retained in the Hamiltonian. Further details are presented in Appendix~\ref{app:1}.
\par
The spatial phase dependence of the pump lasers imprints a site-dependent tunneling phase $\gamma_m=m\gamma=m k_L/(2\pi/d_y)$. Hence, hopping around a plaquette, the wave-function acquires a total phase $\phi=2\pi \gamma$, which can be related to an electron moving in a periodic potential threaded by a magnetic field of strength $|B|=2 \pi \gamma /(d_y^2 e)$. 
\par 
The effective magnetic field breaks the translation symmetry of the original lattice and the Hamiltonian is invariant under a combination of discrete translation and a gauge transformation, i.e., magnetic translation. In particular, when $\gamma=p/q$ is a rational number with $p$ and $q$ being two integers, and 
the energy spectrum splits into $q$ sub-bands, which cluster in a highly fractal structure known as Hofstadter butterfly~\cite{hofstadter1976energy}. 
\par 
In contrast to free space setups the hopping amplitude in the cavity-direction depends on the cavity field amplitude and the effective magnetic field appears only for non-zero cavity-field. Here the coherent amplitude $\langle a \rangle =\alpha$ is determined by the steady-state solution of the mean-field equation:  
\begin{equation}
   \frac{ \partial\alpha}{\partial t}=-(\Delta_c-i\kappa)\alpha-\eta\Theta=0,
\end{equation}
where
\begin{equation}
\Theta=\sum_{l,m}\left(e^{-2i\pi \gamma m} \langle f^\dagger_{l,m} f_{l-1,m}\rangle+e^{2i\pi\gamma m} \langle f^\dagger_{l,m}f_{l+1,m}\rangle\right)
\end{equation}
is the atomic order parameter, which reveals emergent currents of
equal number of left and right moving atoms along the
cavity axis. The order parameter $\Theta$ needs to be self-consistently determined by diagonalizing the Hamiltonian at fixed amplitude $\alpha$,
\begin{equation}
\Theta=\frac{2}{ N_k^2}  \sum_m \sum_{s=1}^q \sum_{\mathbf k\in \mathrm{B.Z.}} n_F(\epsilon_{s,\mathbf k})\cos(2\pi m \gamma) |v_{s,\mathbf k} (m)|^2.
\label{photoneq}%
\end{equation}
Here $\epsilon_{s,\mathbf k}$ and $v_{s,\mathbf k} (m)$ are the eigenvalues and eigenstates of the Harper equation~\cite{harper1955single} 
\begin{align}  \label{harpereq}%
J_y[e^{i k_y}w_{\mathbf k} (m+1)+e^{-i k_y}w_{\mathbf k} (m-1)]&+\nonumber\\
 2\eta (\alpha+\alpha^*)\cos(k_x -2\pi m \gamma )  w_{\mathbf k} (m)
 &=\epsilon w_{\mathbf k}(m).
\end{align}
We use the following Ansatz for the atomic wave-function $\Psi(l,m)=e^{i k_x l}e^{i k_y m}w_\mathbf{k}(m)$,  with $w_\mathbf{k}=\sum c_s v_{s,\mathbf{k}}(m)$ a linear superposition of the eigenstates of the Hamiltonian.
\par
Equations~\eqref{photoneq} and~\eqref{harpereq} are solved self-consistently within the reduced Brillouin zone $k_x\in [-\pi,\pi]$ and  $k_y\in [-\pi/q,\pi/q]$, for a magnetic unit cell with periodic boundary conditions in $x$ and $y$ directions. We focus on the contribution of the bulk to the superradiance, neglecting boundary effects which appear in a pair of chiral edge states~\cite{sheikhan2016cavity}.
\begin{figure}[t]
\centering
\includegraphics[width=0.43\textwidth]{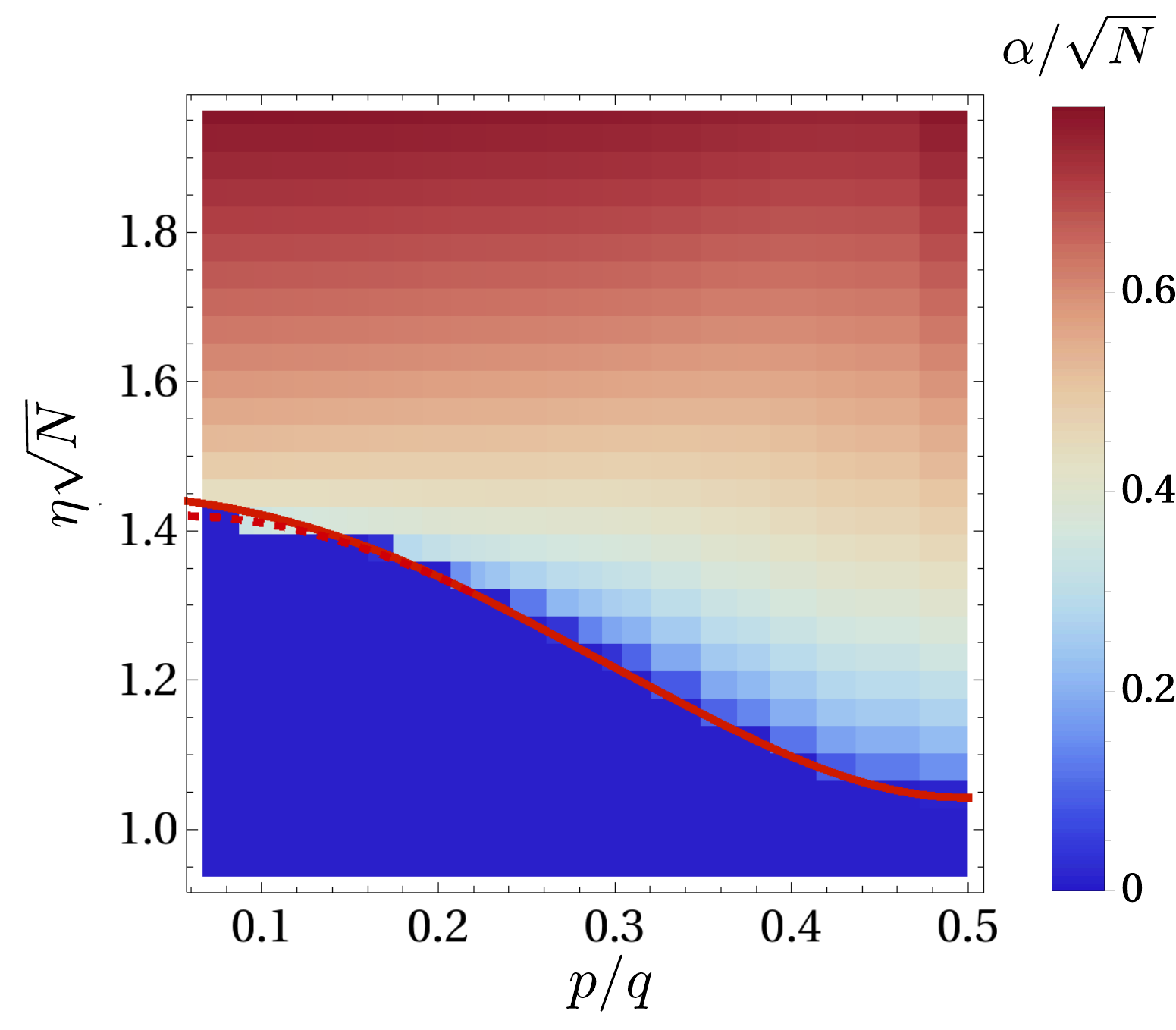}
\caption{Phase boundary (red line) as function of effective flux $\gamma/2\pi=p/q$ and  rescaled pump field $\eta\sqrt{N}$ using the field amplitude modulus $|\alpha|/\sqrt{N}$ as background color.  Note that $p/q$ is discrete and rational, with $1<p<7$ and $1<q<15$. The field amplitude is determined selfconsistently for a Fermi gas at half-filling at fixed finite temperature $k_BT=0.5 E_R$, where $E_R=\hbar^2 k_c^2/2m$ is the recoil energy.  At small fluxes, $\gamma<0.21$, the system exhibits a first-order phase transition, while for bigger fluxes it is of second order. The solid red line shows the analytical result for the critical threshold and the red dashed line the beginning of the region of hysteresis.}
\label{fig:2}
\end{figure}
\begin{figure}[t]
\centering
\includegraphics[width=0.4\textwidth]{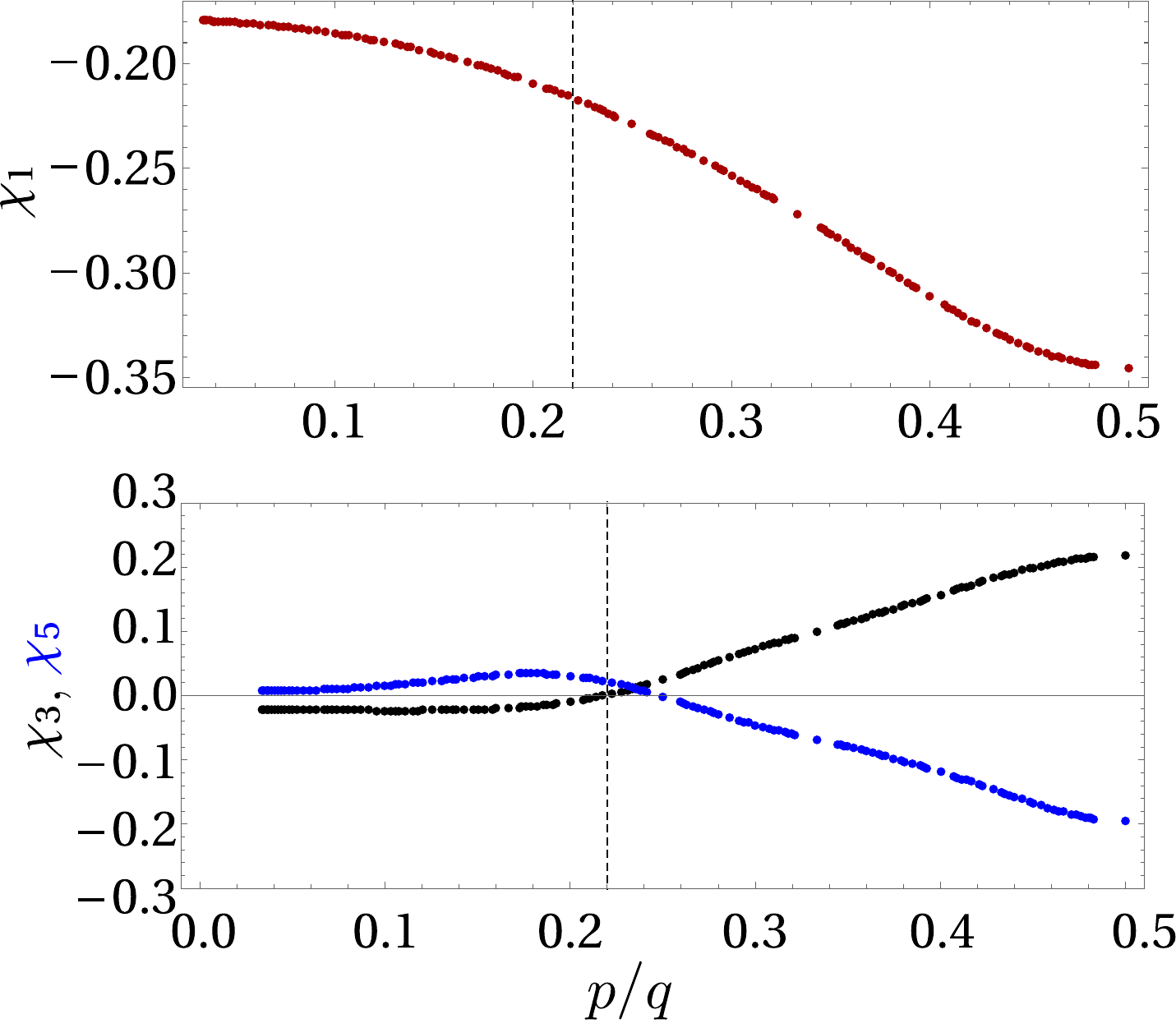}
\caption{Atomic susceptibilities, $\chi_1$ (red), $\chi_3$ (black) and $\chi_5$ (blue) at $k_bT=0.5E_R$. The third order susceptibility $\chi_3$ becomes negative below $p/q=0.21$, signaled by the dashed black line.} 
\label{fig:3}
\end{figure}
\section{\label{sec:3}Results}

\subsection{Phase diagram}
For weak pump $\eta \sqrt{N}$ the system is in the uncoupled normal state (N), i.e., the atoms form a collection of independent chains in the $y$-direction and the cavity is empty. Increasing the effective pump strength the system exhibits a transition to a superradiant (SR) state, where photons are resonantly scattered into the cavity mode and the hopping in cavity ($x$)-direction builds up. 

\par
The stationary cavity-field amplitude is depicted in Fig.~\ref{fig:2}. It grows continuously above the superradiant threshold for large magnetic flux ($0.21<\gamma<0.5$) but displays a non-continuous jump at lower $\gamma < 0.21$. In order to better understand the change from a second to a first order phase transition, as presented in Appendix~\ref{app:2}, we expand the free energy of the system in the Landau form up to sixth order in the atomic order parameter:
\begin{subequations}\label{second:main3}
\begin{align}
F\sim & (1-\frac{4\Delta_c}{\Delta_c^2+\kappa^2}\chi_1\eta^2)|\Theta|^2-\frac{8\Delta_c^3}{(\Delta_c^2+\kappa^2)^3}\chi_3\eta^6|\Theta|^4\tag{\ref{second:main3}}\\ &-\frac{64\Delta_c^5}{3(\Delta_c^2+\kappa^2)^5}\chi_5 \eta^{10}|\Theta|^6\nonumber.
\end{align}
\label{freeenergy}%
\end{subequations}

\par
The effective optical response of the Fermi gas after cycles of absorption and emission of cavity photons is determined by the static susceptibilities, $\chi_i$ (Fig.~\ref{fig:3}). The linear susceptibility $\chi_1$ determines the phase transition threshold 
\begin{equation}
\sqrt{N}\eta_c=\sqrt{\frac{\Delta_c^2+\kappa^2}{4\Delta_c\chi_1}N},
\end{equation} 
which is shown as a red solid line in Fig.~\ref{fig:2}. The sign of  $\chi_3$ determines the order of the phase transition.
\par
In particular, for strong magnetic fields we have $\chi_3 > 0$ and the transition is of the second order. The atoms then behave like a Kerr medium~\cite{carusotto2013quantum}, inducing an intensity dependent shift of the refractive index, $n=n_0+n_2 I$, with $n_2= -8\chi_3\eta^2\Delta_c^3(\Delta_c^2+\kappa^2)$. For decreasing magnetic field the third order susceptibility monotonically decreases becoming negative at $\gamma\simeq 0.21$, which renders the transition first order (bottom panel of Fig.~\ref{fig:3}). In this regime higher order susceptibilities only slightly depend on the magnetic flux $\gamma$. In fact the atomic orbit size significantly exceeds the unit cell of the original lattice, making the lattice structure negligible. The system then exhibits a universal behaviour and the band structure corresponds to Landau levels in free space. 
\par
\begin{figure}[t]
\centering
\includegraphics[width=0.5\textwidth]{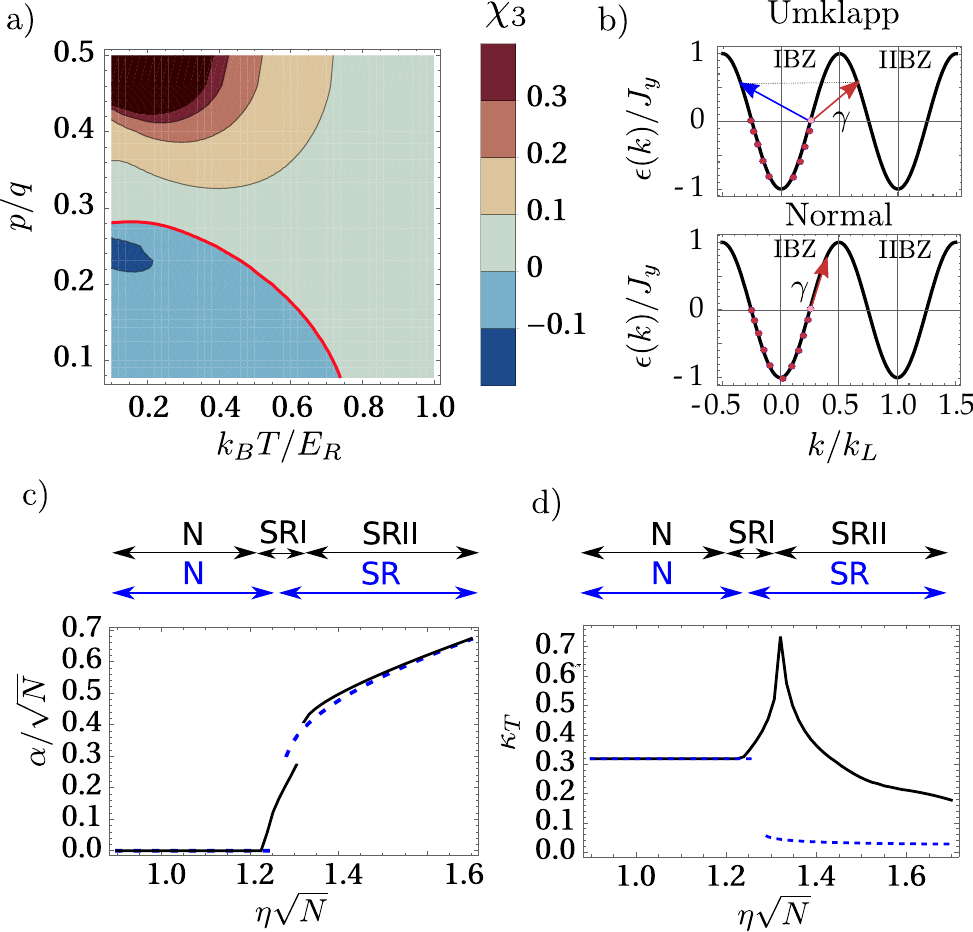}
\caption{(a) Third order susceptibility, $\chi_3$, as a function of the temperature and effective magnetic flux, $2\pi\gamma=2\pi p/q$, with $1<p<6$ and $1<q<13$. The red line corresponds to  zero susceptibility, separating positive and negative regions. (b) An atom at the Fermi surface is scattered after absorbing a photon to a higher energy state, via an Umklapp (top panel) or a normal process (bottom panel). The process is depicted using two Brilliouin zones of the original lattice. Cavity field amplitude (c) and isothermal compressibility (d) at $k_bT=0.05E_R$, for $\gamma=1/3$ (solid black) and $\gamma=1/4$ (dashed blue).} 
\label{fig:4}
\end{figure}
\subsection{First-order transition}

 At low $\gamma$ the emergent magnetic field has only little influence on the system dynamics.  The temperature and the presence of an open Fermi surface then play a fundamental role in order to unravel the physical origin of the first order behaviour of the phase transition. By inspection of the temperature dependence of $\chi_3$ for a Fermi gas at half-filling, we can identify an important change around $\gamma\approx k_F/k_L= 1/4$ (Fig.~\ref{fig:4}(a)). The susceptibility $\chi_3$ is either positive at any temperature, or becomes negative at low temperature. The two regions are separated by the red solid line in Fig.~\ref{fig:4}(a).
\par In the latter case the phase transition becomes first order at low temperatures. This coincides with the regime where scattering one photon keeps the atomic momentum state within the same first Brillouin zone of the original lattice (normal scattering). In contrast, the transition becomes second order when the photon scattering is an Umklapp process (Fig.~\ref{fig:4}(b)), i.e., by inverting the direction of the atomic motion, a momentum transfer ($G=nk_L$) to the optical lattice is required. However, the occupation of higher energy states at higher temperature can favour the Umklapp processes at the expense of direct scattering enhancing the rate to scatter to the next Brillouin zone even for a small momentum transfer. This explains why at higher temperature a second order phase transition occurs and the critical temperature at which this happens increases for small $\gamma$ (Fig.~\ref{fig:4}(a)).

These results are confirmed by the numerical simulations at lower temperatures, $k_bT=0.05E_R$. The re-scaled cavity amplitude as function of the pump strength either grows continuously around the threshold for $\gamma=1/3$ (black line in Fig.~\ref{fig:4}(c)), or exhibits a jump at the critical point for $\gamma=1/4$ (blue dashed line in Fig.~\ref{fig:4}(c)). For $\gamma=1/3$ the rescaled amplitude shows an additional jump at higher pumps $\eta>\eta_c$, hinting that an additional first order transition inside the superradiant phase can appear. Such transition occurs when the cavity-induced hopping exceeds the hopping in the $y$-direction, $J_x/J_y=\eta(\alpha+\alpha^*)=1$. The two superradiant states are characterized by the same order parameter but different isothermal compressibility, $\kappa_T=(1/\rho^2) \partial \rho /\partial \mu$, where $\rho$ is the density of the Fermi gas. This divides the superradiant region into two phase zones: SRI and SRII. In many respects this suggests a liquid-gas type of transition between the SRI and SRII phases, as confirmed by the rapid growth of density fluctuations that can be inferred from the divergence of the compressibility at the critical point (Fig.~\ref{fig:4}(d)). The transition is reminiscent of the case  observed for fermions in linear cavities without external optical lattice~\cite{keeling2014fermionic}. In the latter case, however, the transition was driven by the coupling to an additional degree of freedom, in a process similar to the Larkin-Pimkin mechanism~\cite{larkin1969phase}.

\begin{figure}[t]
\centering
\includegraphics[width=0.45\textwidth]{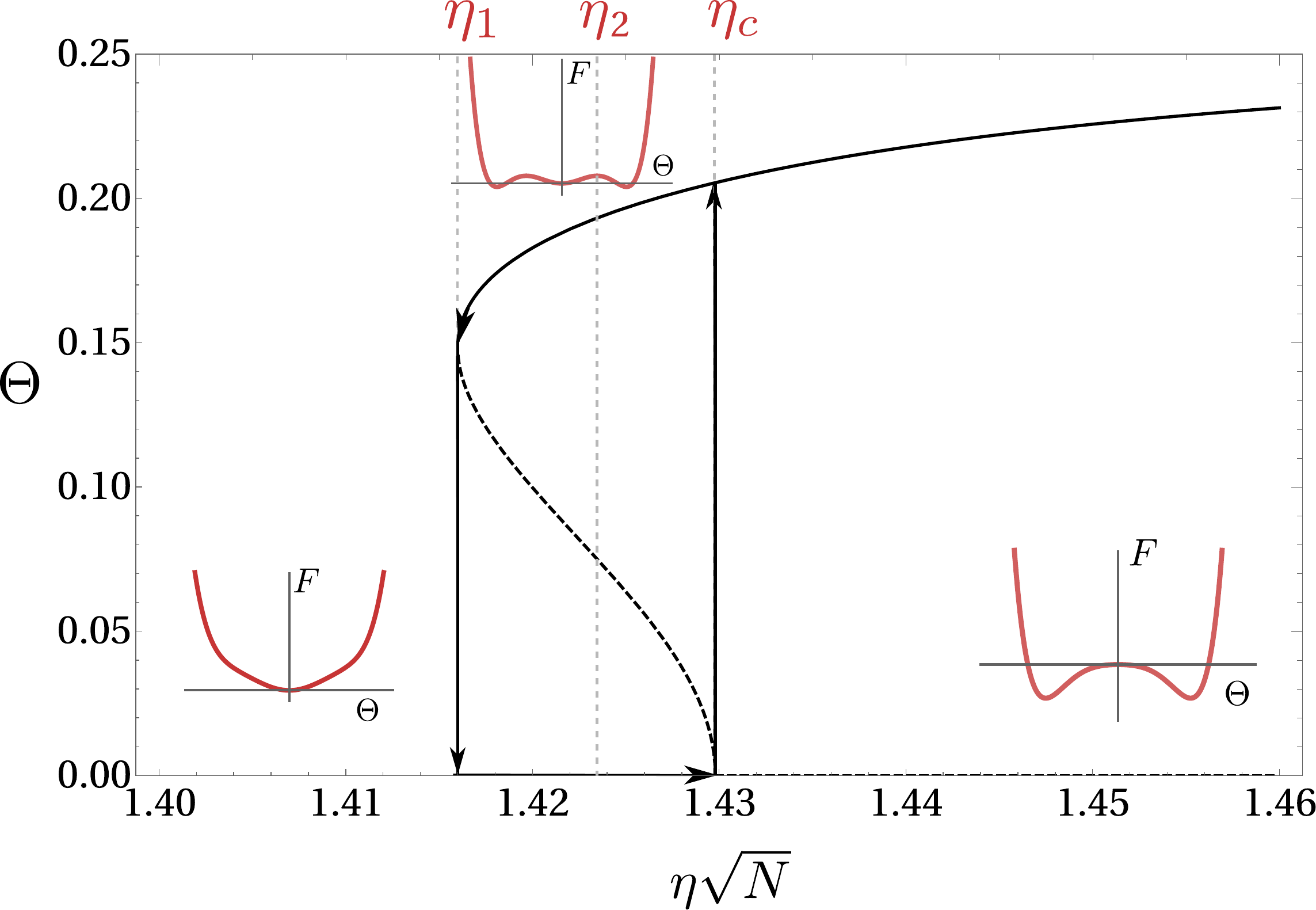}
\caption{Atomic order parameter at $T=0.5E_R$ for $\gamma=1/12$ as a function of the effective pump $\eta\sqrt{N}$. The arrows shows the hysteresis loop and the dotted line represent the metastable solution. The insets show a qualitative picture of the free energy in the different regimes. }
\label{fig:5}
\end{figure}

\subsection{Hysteresis}
For small magnetic flux the system exhibits a bi-stable hysteresis behaviour near the superradiant threshold $\eta_c$. The hysteresis loop and a qualitative picture of the free energy in the different regions are shown in Fig.~\ref{fig:5}.
 As can be seen in the insets, below the threshold
\begin{equation}
 \eta_1=   \frac{\eta_c}{\sqrt{1-\chi_3^2/(12\chi_1\chi_5)}},
\end{equation}
the solution with $\alpha=0$ (empty cavity) is the only minimum of the free energy. Between $\eta_1<\eta<\eta_c$ the free energy has three minima, either local or absolute. The solution for $\alpha\neq0$ is metastable for $\eta_1<\eta<\eta_2$, with
\begin{equation}
    \eta_2=\frac{\eta_c}{\sqrt{1-3\chi_3^2/(8\chi_1\chi_5)}}.
\end{equation}
Between $\eta_2<\eta<\eta_c$, the zero field solution $\alpha=0$ is metastable and finally ceases to be a minimum at $\eta_c$, where the system becomes superradiant.\\

\subsection{\label{subsec:2}Dynamical Hofstadter Butterfly}
\begin{figure}[t]
\centering
\includegraphics[width=0.5\textwidth]{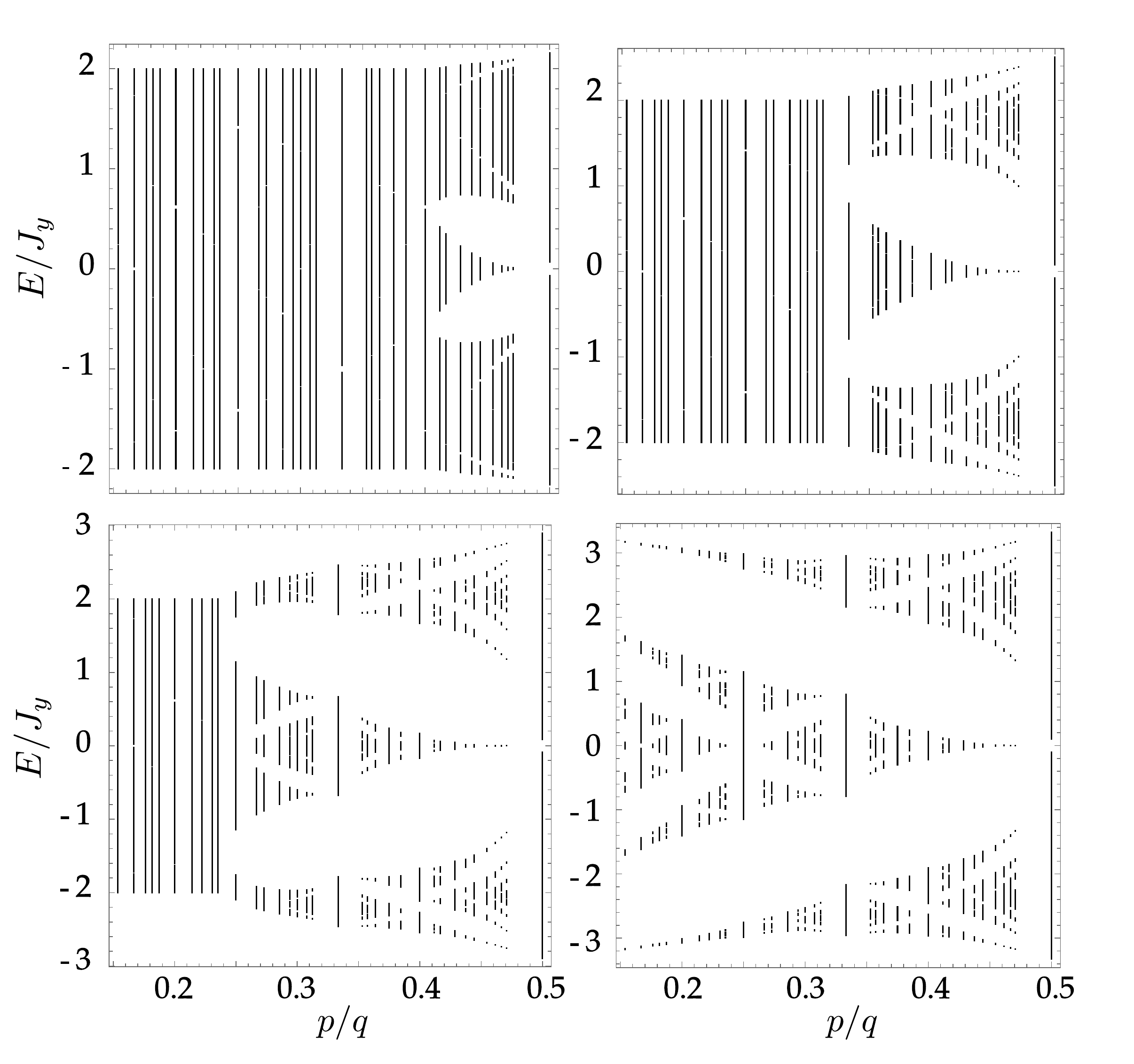}
\caption{Energy spectrum as function of flux $p/q$ for four different pump strength $\eta\sqrt{N}=\{1.1,1.2,1.3,1.4\}E_R$ from top left to bottom right corner at $k_bT=0.5 E_R$. The spectrum initially shows singular shapes and reduces to the conventional Hofstadter butterfly at strong pump.}
\label{fig:6}
\end{figure}

Figure~\ref{fig:6} shows the energy spectrum as a function of the magnetic flux $p/q$ for increasing pump strength $\eta\sqrt{N}$. The magnetic field, $B\sim p/q $, emerges spontaneously with the cavity field amplitude and leads to the opening of $q-1$ gaps in the band structure. As the superradiant phase is entered already at lower pump power for stronger magnetic field, the gap opening progressively extends toward $p/q=0$ as the pump is increased.  
\par
The different structures visible in the energy spectrum strongly depend on the pump strength. At low pump strength (top panels of Fig.~\ref{fig:6}) the gaps organize in the shape of a small butterfly confined in the region of large magnetic fields $0.21<\gamma<0.5$. The gaps close at the boundary of this region, where the amplitude of the cavity field is infinitesimally small. 
When the pump is increased the Hofstadter butterfly is entirely retrieved (right-bottom panel in Fig.~\ref{fig:6}) like in a static optical lattice. The gaps will gradually close, generating a 1D tight-binding in the $x$-direction with bandwidth, $2J_x=2\eta(\alpha+\alpha^*)$. In fact, the system evolves toward a regime of very weakly coupled 1D chains in the $x$-direction, for which the magnetic field can be gauged out. 
\par
The distortion of the energy spectrum, compared to the conventional Hofstadter butterfly~\cite{hofstadter1976energy}, is due to the dynamical nature of the coupling between atoms and cavity photons. At a fixed magnetic field, the system spontaneously chooses the most favourable amplitude of the cavity field, i.e, the effective hopping parameter, $J_x=\eta(\alpha+\alpha^*)$. As the system becomes superradiant the effective Lorentz force exerted by the artificial magnetic field favours the tunneling in the $x$-direction, resulting in an asymmetry of the tunneling amplitudes. Therefore, the energy spectrum can be seen as the superposition of different Hofstadter butterflies with asymmetric hopping, $J_x-J_y$. While the fractal structure is preserved by the form of the Hamiltonian as the hopping phase is not cavity-dependent, the size of the gaps are set by the ratio of the hopping parameters and are characterized by a non-trivial dependence on the magnetic flux $2\pi p/q$.  
\par
\begin{figure}[t]
\centering
\includegraphics[width=0.45\textwidth]{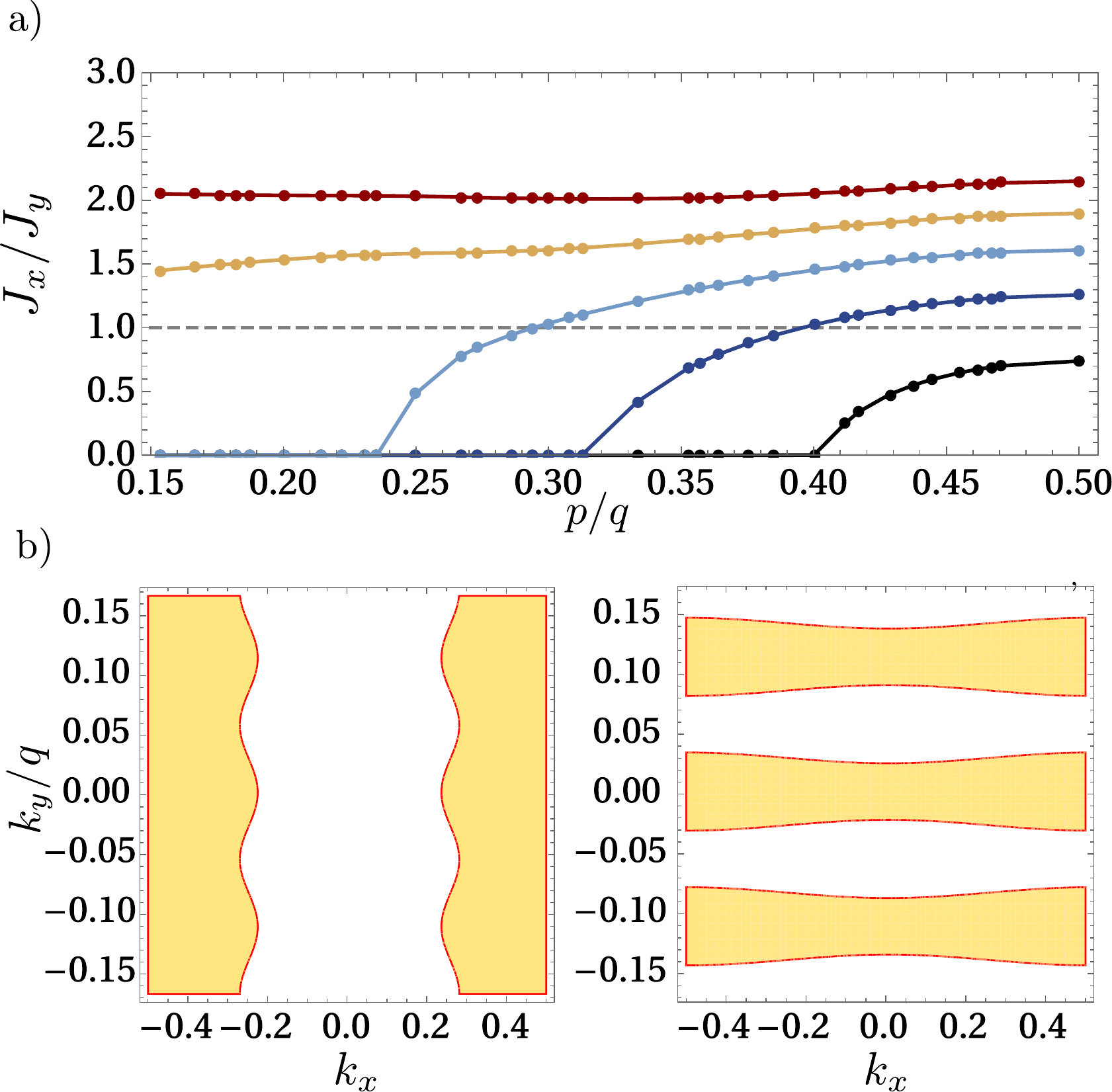}
\caption{ (a) Effective cavity induced hopping as a function of flux $p/q$ at different pumping strengths. Parameters: $\eta\sqrt{N}=\{1.1,1.2,1.3,1.4,1.5\}E_R$ in black, dark blue, light blue, yellow and red respectively. (b) Fermi surface at $\gamma=1/3$ for $k_bT=0.5E_R$ for $\eta\sqrt{N}=1.2E_R$ (left) and $\eta\sqrt{N}=1.3E_R$ (right).}
\label{fig:7}
\end{figure}

 This is illustrated in Fig.~\ref{fig:7}(a), where the hopping ratio $J_x/J_y$ is shown as a function of the magnetic flux for different pump strengths. In the weak pump regime (black and dark blue lines) the dynamic butterfly is a superposition of static Hofstadter butterflies with very different effective hopping amplitudes. The hopping in the $x$-direction grows as the magnetic field is increased but remains rather small compared to the hopping in the other direction. As a consequence the curvature of the band structure and the Fermi surface align along $y$-direction, see left panel in Fig.\ref{fig:7}(b). 
 
As the pump is increased, the field amplitude and the hopping in the $x$-direction become almost independent of the magnetic flux (red and yellow line in Fig~\ref{fig:7}(a)). In this regime the kinetic energy in the $x$-direction dominates and the Fermi surface aligns along the cavity axis. Note that at low temperature this is accompanied by the onset of a first order transition within the superradiant phase, SRI-SRII, as shown in the previous section. 
\section{\label{sec:4}Conclusions and Outlook}
 We have shown that non-linear coupling between atomic motion and a cavity field mode offers a new perspective on the generation of synthetic dynamical magnetic fields. In contrast to free space, the gauge field emerges spontaneously via maximizing the light scattered into the cavity and changing the atomic density configuration. The complex interplay between the fractal structure of the energy bands and the superradiant scattering thus generates new shapes for a dynamical Hofstadter butterfly.    
\par
Note that atoms are coupled only to a specific wave-length of the light field determined by the chosen cavity mode. As shown recently employing several distinct cavity modes the system gets more freedom and a global symmetry can ``emerge'' in a cavity-QED system~\cite{mivehvar2019emergent}. Therefore, generalization of our studied system to multi-mode cavities and in particular a ring or fiber geometry~\cite{holzmann2018synthesizing} could allow to fully reproduce the minimal coupling of a charged particle to a local $U(1)$ gauge potential. Making use of the dynamical coupling between light and atoms in cavity systems is a promising route toward the experimental realization of synthetic dynamical gauge fields. Moreover, on a different level, the mediation of long-range two-body interactions due to the exchange of photons can lead to the observation of exotic states, as particles with anyonic statistics in fractional quantum Hall states.

\section*{Acknowledgments}
 F.~M.\ is grateful to Nathan Goldman for fruitful discussions. F.~M.\ is supported by the Lise-Meitner Fellowship M2438-NBL of the Austrian Science Fund (FWF), and the International Joint Project
No.\ I3964-N27 of the FWF and the National Agency for
Research (ANR) of France.

\appendix

\section{\label{app:1}Effective Hamiltonian}

\begin{figure}[b]
\centering
\includegraphics [width=0.5\textwidth]{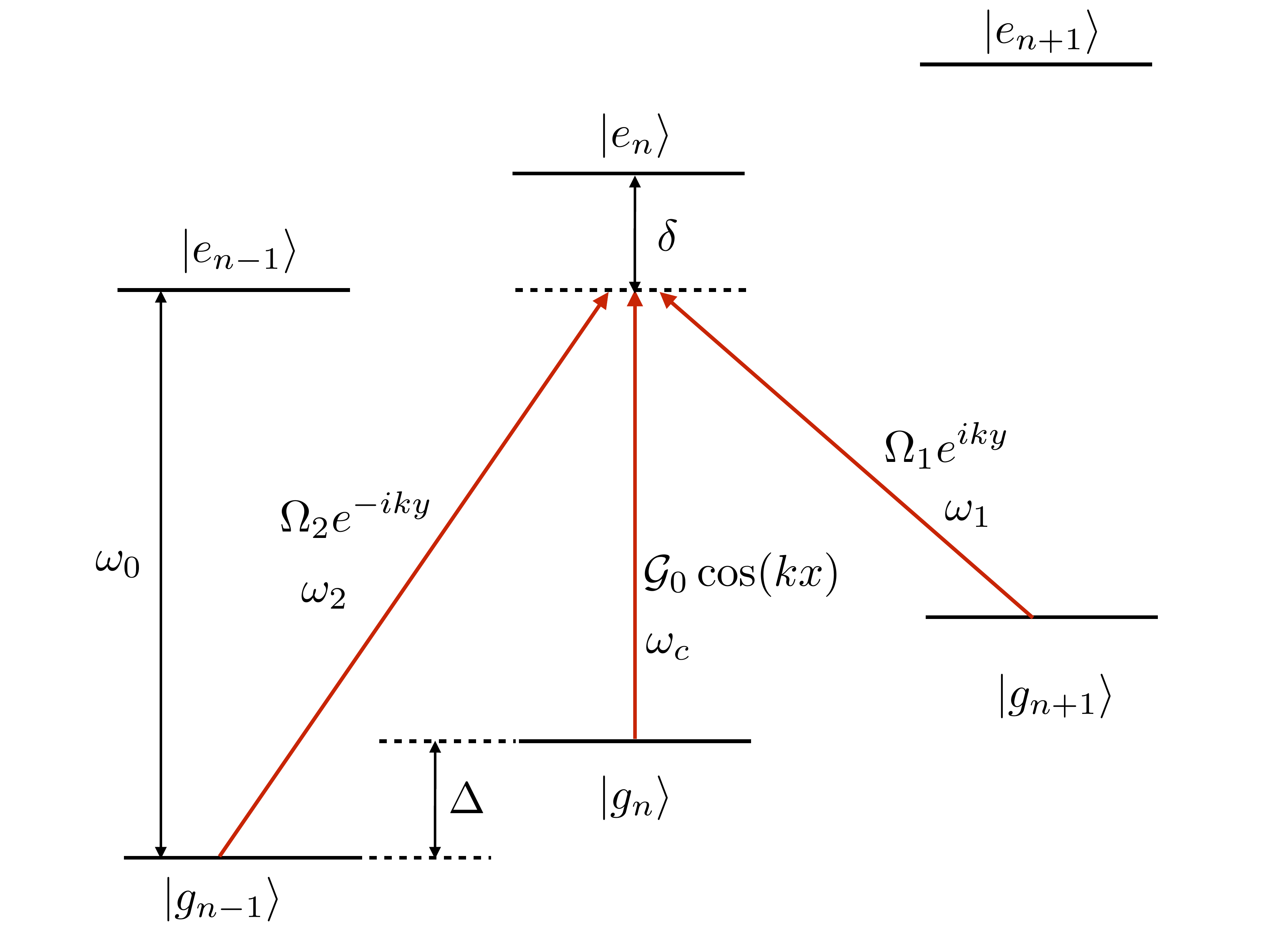}
\caption{Three generic lattice sites along $x$ direction.}
\label{fig:three-state-manifold}
\end{figure}

Consider atoms loaded into a 2D optical lattice of lattice constant, $\mathbf d=[d_x,d_y]$. The hopping along $x$-direction is at first suppressed due to the potential offset $\Delta$ between adjacent lattice sites and then restored thanks to the cavity- and laser-assisted hoppings. The hopping along $y$-direction is due to the kinetic energy of the atoms. Let us just focus in the $x$-direction and consider three generic lattice sites labeled $n-1$, $n$, and $n$ as in Fig.~\ref{fig:three-state-manifold}. First consider only transitions which involves the atomic excited state in site $n$, that is, $\ket{e_n}$. The Hamiltonian $H=H_0+H_{\rm int}$ reads ($\hbar=1$),
\begin{align}
H_0=-(\omega_0+\Delta)\sigma_{n-1}-\omega_0\sigma_{n}-(\omega_0-\Delta)\sigma_{n+1}
+\omega_ca^\dag a,
\end{align}
\begin{align}
H_{\rm int}&=\Omega_2 e^{-iky} e^{-i\omega_2t} \sigma_{n-1}^++g_0\cos{(kx_n)}a\sigma_{n}^+\nonumber\\
&+\Omega_1 e^{iky} e^{-i\omega_1t} \sigma_{n+1}^++\text{H.c.},
\end{align}
where $\sigma_{n-1}=\ket{g_{n-1}}\bra{g_{n-1}}$, $\sigma_n=\ket{g_{n}}\bra{g_{n}}$ , $\sigma_{n+1}=\ket{g_{n+1}}\bra{g_{n+1}}$, $\sigma_{n-1}^+=\ket{e_{n}}\bra{g_{n-1}}$, $\sigma_{n}^+=\ket{e_{n}}\bra{g_n}$, $\sigma_{n+1}^+=\ket{e_{n}}\bra{g_{n+1}}$. For simplicity a two-photon resonance is assumed $\omega_c=\omega_1+\Delta=\omega_2-\Delta$ in the following and $k\equiv k_c\simeq k_1\simeq k_2$.

Applying the unitary transformation $U=\exp{\{-i[\omega_2\sigma_{n-1}+\omega_p(\sigma_{n}-a^\dag a)+\omega_1\sigma_{n+1}]t}\}$ to the Hamiltonian $H$ yields,
\begin{align} \label{eq:rotating-frame-H}
\tilde{H}&=\delta(\sigma_{n-1}+\sigma_{n}+\sigma_{n+1})
+[\Omega_2 e^{-iky} \sigma_{n-1}^+\nonumber\\&+g_0\cos{(kx_n)}a\sigma_{n}^+
+\Omega_1 e^{iky} \sigma_{n+1}^++\text{H.c.} ],
\end{align}
where $\delta=\omega_c-\omega_0\sim  \omega_p-\omega_0$, with $\omega_p=(\omega_1+\omega_2)/2$ the average pump frequency. Here we have made use of the relations $U \sigma_{n-1}^+U^\dag= e^{i\omega_2t}\sigma_{n-1}^+$ etc. and $\tilde{H}=UHU^\dag+i(\partial_t U)U^\dag.$ We find the stationary values of the operators $\sigma_{n-1}^+$, $\sigma_{n}^+$, $\sigma_{n+1}^+$ by setting to zero the the Heisenberg equation of motion $i\partial_t O=[O,\tilde{H}]$ upon assuming a large detuning $\delta$
\begin{align} \label{eq:ss-operators}
\sigma_{n-1}^+&\simeq\frac{1}{\delta}
(\Omega_2^* e^{iky} \sigma_{n-1}+\Omega_1^* e^{-iky} \sigma_{n+1,n-1}
\nonumber\\&+g_0\cos{(kx_n)}a^\dag\sigma_{n,n-1}),\nonumber\\
\sigma_n^+&\simeq\frac{1}{\delta}
(\Omega_2^* e^{iky} \sigma_{n-1,n}+\Omega_1^* e^{-iky} \sigma_{n+1,n}
\nonumber\\&+g_0\cos{(kx_n)}a^\dag\sigma_{n}),\nonumber\\
\sigma_{n+1}^+&\simeq\frac{1}{\delta}
(\Omega_2^* e^{iky} \sigma_{n-1,n+1}+\Omega_1^* e^{-iky} \sigma_{n+1}
\nonumber\\&+g_0\cos{(kx_n)}a^\dag\sigma_{n,n+1}),
\end{align}
where $\sigma_{n,n-1}=\ket{g_{n}}\bra{g_{n-1}}$, $\sigma_{n+1,n-1}=\ket{g_{n+1}}\bra{g_{n-1}}$, $\sigma_{n+1,n}=\ket{g_{n+1}}\bra{g_{n}}$, etc. Here we have also assumed a negligible population of the excited state, $\ket{e_{n}}\bra{e_{n}}\simeq0$, due to the large detuning $\delta$.

Substituting Eq.~\eqref{eq:ss-operators} back in the Hamiltonian~\eqref{eq:rotating-frame-H} yields the effective Hamiltonian,  
\begin{align} \label{eq:n-effective-H}
\tilde{H}_{\rm eff}^{(n)}&=\frac{2}{\delta}\{
g_0^2\cos^2{(kx_n)}a^\dag a\sigma_{n}\nonumber\\
&+[\Omega_2g_0 e^{-iky}\cos{(kx_n)}a^\dag\sigma_{n,n-1}
\nonumber\\&+\Omega_1^*g_0 e^{-iky}\cos{(kx_n)}a\sigma_{n+1,n}
+\text{H.c.}]\},
\end{align}
where the constant terms proportional to $\Omega_1$ and $\Omega_2$, and terms involving next nearest neighbour scattering $\sigma_{n+1,n-1}$ have been omitted.

Considering now transitions which involve the states $\ket{e_{n\pm1}}$ results in the following contributions to the $\{n-1,n,n+1\}$ manifold,
\begin{align} \label{eq:npm1-effective-H}
\tilde{H}_{\rm eff}^{(n-1)}&\propto\frac{2}{\delta}\{
g_0^2\cos^2{(kx_{n-1})}a^\dag a\sigma_{n-1}\nonumber\\
&+[\Omega_1^*g_0 e^{-iky}\cos{(kx_{n-1})}a\sigma_{n,n-1}
+\text{H.c.}]\},\nonumber\\
\tilde{H}_{\rm eff}^{(n+1)}&\propto\frac{2}{\delta}\{
g_0^2\cos^2{(kx_{n+1})}a^\dag a\sigma_{n+1}\nonumber\\
&+\left[\Omega_2g_0 e^{-iky}\cos{(kx_{n+1})}a^\dag\sigma_{n
1,n}
+\text{H.c.}\right]\}.
\end{align}
Assuming $\Omega_1=\Omega_2=\Omega\in\mathbf{R}$ and $\lambda_c=2\pi/k=d_x$, the total effective Hamiltonian takes the form,
\begin{align} \label{eq:effective-H}
\tilde{H}_{\rm eff}&=\frac{2}{\delta} \sum_{n}\{
g_0^2\cos^2{(kx_n)}a^\dag a\sigma_{n}
\nonumber\\ &+\Omega g_0(a+a^\dag)\left[e^{-iky}\cos{(kx_n)}\sigma_{n,n-1}
+\text{H.c.}\right]\},
\end{align}
or in the second-quantized tight-binding formalism 
\begin{align} \label{eq:effective-H-2nd-TB}
\tilde{H}_{\rm eff}&=a^\dag a\sum_{n,m}\epsilon_{n,m}c^\dag_{n,m}c_{n,m}
\nonumber\\&+(a+a^\dag)\sum_{n,m}\left(J_{n,m}^xe^{-iky_m}c^\dag_{n,m}c_{n-1,m}+\text{H.c.}\right)\nonumber\\
&+J^y\sum_{n,m}\left(c^\dag_{n,m}c_{n,m-1}+\text{H.c.}\right),
\end{align}
where the hopping along the $y$ direction is now also included. The matrix elements are given by,
\begin{align} 
\epsilon_{n,m}&=\frac{2}{\delta}g_0^2
\int\int dx dy \cos^2{(kx)} \nonumber\\
& \times |W(x-x_n)W(y-y_n)|^2
\nonumber\\&=\frac{2}{\delta} g_0^2
\int\int dx \cos^2{(kx)} |W(x-x_n)|^2,\nonumber\\
J_{n,m}^xe^{-iky_m}&=\frac{2}{\delta}\Omega g_0
\int\int dx dy W^*(x-x_n)W^*(y-y_n)\nonumber\\&\times e^{-iky}\cos{(kx)} W(x-x_{n-1})W(y-y_m)\nonumber\\
&=\frac{2}{\delta}\Omega g_0
\int dx \cos{(kx)}\nonumber\\
&\times  W^*(x-x_n) W(x-x_{n-1})\nonumber\\
&\times \int dy e^{-iky} W^*(y-y_m) W(y-y_m),
\end{align}
where $W(\mathbf X- \mathbf{R})=W(x-x_n)W(y-y_m)$ is the ground state Wannier function describing particles localized at the site $[n,m]$.

\section{\label{app:2}Free energy expansion}
In order to derive an effective Landau theory for the atomic order parameter $\Theta$, as defined in the main text, we start from the the action of the system expressed in momentum space 
\begin{subequations}
\begin{align}
S[\alpha,\alpha^*,&c^\dagger_{k_x,k_y},c_{k_x,k_y}]=\Delta_c|\alpha|^2\\ \nonumber&+\frac{1}{\beta V}\sum_{n,k_x,k_y}\left( i\omega_n-2J_y\cos(k_y)\right) c^\dagger_{n,k_x,k_y}c_{n,k_x,k_y}\\ \nonumber&-\eta(\alpha+\alpha^*)\frac{1}{\beta V} \sum_{k_x,k_y} \Big(e^{-k_x}c^\dagger_{k_x,k_y}c_{k_x,k_y+\gamma}\\ \nonumber&+e^{-k_x}c^\dagger_{k_x,k_y}c_{k_x,k_y-\gamma}\Big).
\end{align}
\end{subequations}
Note that only the static component of the bosonic field $\alpha$ is retained, which is linearly related to the atomic order parameter by the equation of motion $\alpha=-\eta \Theta/(\Delta_c-i\kappa)$. We integrate out fermionic degrees of freedom, obtaining an effective action for the photonic field only, $S_{eff}[\alpha,\alpha^*]=\Delta_c|\alpha|^2+\rm{tr}\enskip \rm{ln} \hat G^{-1}$. The trace operator
\begin{equation}
\rm{tr}\enskip \rm{ln} \hat G^{-1}= \rm{tr}\enskip \rm{ln} G_0^{-1}-\sum_n \frac{1}{2n}tr(G_0\Gamma)^{2n}
\end{equation}
is obtained by perturbatively expanding the Green function $G(\mathbf k,i\omega_n)$ around the zero order one
\begin{widetext}
\begin{equation}
G_0^{-1}(\mathbf{k},\omega_n)=\begin{bmatrix}
\ddots&0&0&0&0\\
0&i\omega_n- 2 J_y \cos(k_y-\gamma)&0&0&0\\
0&0&i\omega_n- 2 J_y \cos(k_y-\gamma)&0&0\\
0&0&0&i\omega_n- 2 J_y\cos(k_y-\gamma)&0\\
0&0&0&0&\ddots\\
\end{bmatrix}
\end{equation} 
\end{widetext}
where the perturbative term is given by the interaction matrix 
\begin{equation}
\Gamma(\mathbf{k})=-\eta(\alpha+\alpha^*)\begin{bmatrix}
0&e^{-i k_x}&0&0&0\\
e^{ i k_x}&0&e^{- i k_x}&0&0\\
0&e^{ i k_x}&0&e^{-i k_x}&0\\
0&0&e^{ i k_x}&0&e^{- ik_x}\\
0&0&0&e^{ i k_x}&0\\
\end{bmatrix}
\end{equation} 
Here, $i\omega_n=\pi(2n+1)/\beta$ are fermionic Matsubara frequencies. By keeping up to the sixth order in $\alpha$, the effective free energy is
\begin{align}
F&=\Delta_c|\alpha|^2-\eta^2\chi_1(\alpha+\alpha^*)^2-\frac{\eta^4}{2}\chi_3(\alpha+\alpha^*)^4\nonumber\\&-\frac{\eta^6}{3}\chi_5(\alpha+\alpha^*)^6,
\end{align}
or in powers of the atomic order parameter, $\Theta$, reads
\begin{align}
F&\sim(1-\frac{4\Delta_c}{\Delta_c^2+\kappa^2}\chi_1\eta^2)|\Theta|^2-\frac{8\Delta_c^3}{(\Delta_c^2+\kappa^2)^3}\chi_3\eta^6|\Theta|^4\nonumber\\&-\frac{64\Delta_c^5}{3(\Delta_c^2+\kappa^2)^5}\chi_5 \eta^{10}|\Theta|^6
\end{align}
The free energy depends on the cavity properties and the coupling with the atoms is enclosed inside the susceptibilities
\begin{subequations}
\begin{align}
\chi_1&=\frac{1}{\beta}\sum_{n,k\in B.Z.} G_k(i\omega_n)G_{k+\gamma}(i\omega_n)\\
\chi_3&=\frac{1}{\beta}\sum_{n,k\in B.Z.}[ G_k^2(i\omega_n)G_{k+\gamma}^2(i\omega_n)\\&+2G_{k-\gamma}(i\omega_n)G_k^2(i\omega_n)G_{k+\gamma}(i\omega_n)]\nonumber\\
\chi_5&=\frac{1}{\beta}\sum_{n,k\in B.Z.}[ G_k^3(i\omega_n)G_{k+\gamma}^3(i\omega_n)\\&+3G_{k-\gamma}^2(i\omega_n)G_k^3(i\omega_n)G_{k+\gamma}(i\omega_n)\nonumber\\
&+3G_{k-\gamma}(i\omega_n)G_k^3(i\omega_n)G_{k+\gamma}^2(i\omega_n)\nonumber\\&+3G_{k}(i\omega_n)G_{k+\gamma}^2(i\omega_n)G_{k+2\gamma}^2(i\omega_n)G_{k+3\gamma}(i\omega_n)]\nonumber
\end{align}
\end{subequations}
The susceptibilities shown in the main text are numerically calculated by truncating the summation over the Matsubara frequencies until convergence with fixed chemical potential $\mu=0$, same for the matrices $G_0(\mathbf k,\omega_n)$ and $\Gamma(\mathbf k)$ which are summed in momentum space over the original Brillouin zone $[-\pi/d_x,\pi/d_y]$.
\subsection{Expansion of the susceptibility for low magnetic fluxes} In order have a better understanding of the physics at low magnetic fluxes, we have analytically computed the expressions for the susceptibilities $\chi_1$ and $\chi_3$. The first order susceptibility is
\begin{equation}
\chi_1=\sum_{k\in \rm{B.Z.}}  \frac{n_F\left(\epsilon_{k+\gamma}\right)-n_F\left(\epsilon_{k}\right)}{\epsilon_{k+\gamma}-\epsilon_{k}}, 
\end{equation}
with $\epsilon_k=J_y\cos(k)$, the tight binding energy along the $y$-direction where we set  $\mu=0$ for half filling. We expand $\chi_1$ for small $\gamma$ 
\begin{equation}
\chi_1(\gamma\ll 1)= \sum_{k\in \rm{B.Z.}} \Big[
-\beta n_F(\cos(k))\left[1-n_F(\cos (k))\right]\Big]
\end{equation}
Note that the linear term vanishes and the main contribution to the linear susceptibility is a constant, which is proportional to the compressibility of a 1D chain of fermionic particles in the tight binding regime. As $n_F(\epsilon)$ is the probability that the state $\epsilon$ is occupied, while $1-n_F(\epsilon)$ is the probability that the state $\epsilon$ is not occupied, their product represent the scattering amplitude of a scattering process between two state of the same energy, which at very low temperature is only possible from one side to the other of the Fermi surface. The next contribution to $\chi_1$ is quadratic and this behaviour can also be observed in the plot of the susceptibilty $\chi_1$, see Fig~\ref{fig:2} in the main text. Note that at the zero order, in $\gamma$ we don't see the effect of the magnetic field but rather the temperature, dimensionality and filling play the fundamental role. \par
The third order $\chi_3$ susceptibilty represents the response of the medium to three photon processes, through cycles of multiple emission and absorption. The full analytics expression is

\begin{align}
\chi_3=&\sum_{k\in \rm{B.Z.}}- 2 \frac{n_F(\epsilon_{k+\gamma})-n_F\left(\epsilon_k \right)}{(\epsilon_{k+\gamma}-\epsilon_k)^3} \nonumber\\&
+\frac{n_F'\left(\epsilon_{k+\gamma}\right)-n_ F'\left(\epsilon_k\right)}{(\epsilon_{k+\gamma}-\epsilon_k)^2}\nonumber\\
&+2 \frac{n_F(\epsilon_{k-\gamma})}{(\epsilon_{k-\gamma}-\epsilon_k)^2(\epsilon_{k-\gamma}-\epsilon_{k+\gamma})}\nonumber\\\nonumber&-2\frac{n_F(\epsilon_{k+\gamma})}{(\epsilon_{k+\gamma}-\epsilon_k)^2(\epsilon_{k-\gamma}-\epsilon_{k+\gamma})}\\
&+2\frac{n_F(\epsilon_k)}{(\epsilon_{k-\gamma}-\epsilon_{k})(\epsilon_{k}-\epsilon_{k+\gamma})}\nonumber\\&\times\Big(\frac{1}{\epsilon_k-\epsilon_{k+\gamma}}+\frac{1}{\epsilon_k-\epsilon_{k-\gamma}}\Big)\nonumber\\
&-2\frac{n_F'(\epsilon_k)}{(\epsilon_{k-\gamma}-\epsilon_k)(\epsilon_k-\epsilon_{k+\gamma})}
\end{align}

In a linear cavity photons are in a superposition state of two conterpropagating momenta. The interaction with the cavity photons induces two type of processes. The first two lines refers to cycles of absorption and emission where the scattering processes always involve interactions with the same momentum component of the photon field. The other lines, refer to scattering processes in which a redistribution of photons between the two momentum component are involved. At the lowest order in $\gamma$, the susceptibility $\chi_3$ becomes 

\begin{align}
\chi_3 (\gamma \ll 1)&=\sum_{k\in\mathrm{B.Z.}} \frac{\beta^3}{6} n_f(\epsilon_k)[1-n_f(\epsilon_k)] \nonumber\\& \times\left[1-6n_f(\epsilon_k)\left[1-n_f(\epsilon_k)\right] \right].
\end{align}


\end{document}